\documentclass[iop,revtex4]{emulateapj}

\usepackage{xcolor} 
 \newcommand{\mic}{$\mu$m}
 \newcommand{\msun}{$M_\odot$}

\received{....}
\revised{...}
\accepted{\today}

\shorttitle{The Star Formation Main Sequence in the HST Frontier Fields}
\shortauthors{Santini et al.}

\begin{document}

\title{The Star Formation Main Sequence in the {\it Hubble Space Telescope} Frontier Fields} 

\email{paola.santini@oa-roma.inaf.it}

\author{Paola Santini\altaffilmark{1}, Adriano Fontana\altaffilmark{1}, Marco Castellano\altaffilmark{1}, Marcella Di
  Criscienzo\altaffilmark{1}, Emiliano Merlin\altaffilmark{1}, Ricardo
  Amorin\altaffilmark{2,3},
Fergus Cullen\altaffilmark{4},  
 Emanuele Daddi\altaffilmark{5},  
 Mark Dickinson\altaffilmark{6},  
James S. Dunlop\altaffilmark{4}, 
 Andrea Grazian\altaffilmark{1}, 
Alessandra Lamastra\altaffilmark{1}, 
Ross J. McLure\altaffilmark{4}, 
 Micha{\l}.~J.~Micha{\l}owski\altaffilmark{7,4}, 
Laura Pentericci\altaffilmark{1}, 
Xinwen Shu\altaffilmark{8}}
\altaffiltext{1}{INAF - Osservatorio Astronomico di Roma, via di Frascati 33,
  00078 Monte Porzio Catone, Italy}
\altaffiltext{2}{Cavendish Laboratory, University of Cambridge, 19 J. J. Thomson
 Ave., Cambridge CB3 0HE, UK}
\altaffiltext{3}{Kavli Institute for Cosmology, University of Cambridge, Madingley Road, Cambridge CB3 0HA, UK}
\altaffiltext{4}{Institute for Astronomy, University of Edinburgh, Royal Observatory,
 Edinburgh, EH9 3HJ, U.K}
\altaffiltext{5}{Laboratoire AIM, CEA/DSM-CNRS-Universit{\'e} Paris Diderot, IRFU/Service d'Astrophysique, B\^at.709, CEA-Saclay, 91191 Gif-sur-Yvette Cedex, France}
\altaffiltext{6}{National Optical Astronomy Observatory, Tucson, Arizona 85719}
\altaffiltext{7}{Astronomical Observatory Institute, Faculty of Physics, Adam
Mickiewicz University, ul.~S{\l}oneczna 36, 60-286 Pozna{\'n}, Poland}
\altaffiltext{8}{Department of Physics, Anhui Normal University, Wuhu, Anhui, 241000, China}



\begin{abstract}
  We investigate the relation between star formation rate (SFR) and
  stellar mass (M), i.e. the Main Sequence (MS) relation of
  star-forming galaxies, at $1.3\leq z < 6$ in the first four HST
  Frontier Fields, based on rest-frame UV observations. Gravitational
  lensing combined with deep HST observations allows us to extend the
  analysis of the MS down to 
  $\log M/$\msun$\sim 7.5$ at $z\lesssim 4$ and $\log M/$\msun$\sim 8$
  at higher redshifts, a factor of $\sim$10 below most previous
  results. We perform an accurate simulation to take into account the
  effect of observational uncertainties 
  and correct for the Eddington bias. This step allows us to reliably
  measure the MS and in particular its slope.  While the normalization
  increases with redshift, we fit an unevolving and approximately
  linear slope. We nicely extend to lower masses the results of
  brighter surveys. Thanks to the large dynamic range in
  mass 
  and by making use of the simulation, we analyzed any possible mass
  dependence of the dispersion around the MS. 
  We find tentative evidence that the scatter decreases with
  increasing 
  mass, 
  suggesting larger variety of star formation histories in low mass
  galaxies. This trend agrees with theoretical predictions,
and is explained as either a consequence
  of the smaller number of progenitors of low mass galaxies in a
  hierarchical scenario and/or of the efficient but intermittent
  stellar feedback processes in low mass halos. Finally, we observe an
  increase in the SFR per unit stellar mass with redshift milder than predicted by
  theoretical models,  implying a still incomplete 
    understanding of the processes responsible for galaxy
    growth. 
\end{abstract}



\section{Introduction}\label{sec:intro}

A key step to understanding galaxy evolution is
  estimating galaxy redshifts and physical properties, in particular
  the star formation rate (SFR) and the stellar mass (M). They
  directly probe the process of gas conversion into stars and
  subsequent stellar mass build-up.  The analysis of large,
statistically significant galaxy samples has allowed the establishment
of the existence of a well-defined relation between the SFR and the
stellar mass, called Main Sequence (MS hereafter), a subject of study of
a large number of papers in the literature
(\citealt{brinchmann04,noeske07,elbaz07,daddi07a,santini09,peng10};
see \citealt{speagle14} for a compilation of results and more
references).

The MS has been traditionally parameterized as a power-law of the form
$\log SFR = \alpha \log M + \beta$.  In recent years, however,
many studies have reported evidence that the MS flattens at high
masses ($M\sim 10^{10-11}M_\odot$)
\citep[e.g.][]{magnelli14,whitaker14,schreiber15}, with the turnover
mass increasing with redshift \citep{tasca15,lee15,tomczak16} and at
the same time becoming less evident \citep{whitaker14,schreiber15}.
Recently, the analysis of \cite{dunlop17} (see also
\citealt{koprowski16}) ascribes the observed flattening at
$z\gtrsim 2$ to the inability of correctly recovering the SFR in
deeply obscured high-mass galaxies when rest-frame optical/UV tracers
are adopted (although some of the works above are based on FIR
estimators, this is qualitatively consistent with the weakening of the
mass turnover with redshift).

The existence of the MS is universally
recognized, and there is general consensus of an increasing
normalization with redshift, associated with a higher rate of gas
accretion in the early Universe. However, the details and in
particular the MS slope vary from one study to the other, ranging from
$\sim0.6$ to 1, most strongly depending on the sample selection and
SFR tracer adopted
\citep{santini09,rodighiero14,speagle14}. Nevertheless, the majority
of recent results point towards a roughly linear and unevolving slope
\citep{whitaker14,schreiber15,tasca15,tomczak16}, at least out to
$z\sim 4$.

Physically, the existence itself and the tightness of the MS relation
(scatter of $\sim$0.25--0.4 dex,
\citealt{rodighiero11,speagle14,schreiber15}) suggest similarity in
the gas accretion histories of galaxies.  The flattening at high
masses has been interpreted as due to either the contribution of the
bulge to the stellar mass (while the SFR comes primarily from the
disk, \citealt{schreiber15}) or the onset of quenching processes
\citep{tasca15}.  According to some work
\citep[e.g.][]{renzini15,whitaker15}, it seems to arise from an
incorrect separation of passive and quenched galaxies from
star-forming ones.

While the majority of galaxies occupy the locus of the MS, outliers
are also observed with intense levels of SFR given their stellar
mass. These two populations have been associated with different growth
mechanisms \citep{daddi10,genzel10,elbaz11}: MS galaxies are thought
to grow on long timescales as a consequence of smooth gas accretion
from the Intergalactic Medium (IGM), while MS outliers, also called
starbursts, seem to be triggered by mergers and form stars with high
efficiency, although this view is currently debated
\citep[e.g.][]{narayanan12a,kennicutt12,santini14,mancuso16}. The
latter, being very rare, despite their high level of SFR, seem to
contribute modestly to the cosmic star formation history
\citep{rodighiero11,sargent12,lamastra13}.

The latest deep multi-band surveys allowed us to observe fainter and
fainter galaxies and probe the MS to stellar masses as faint as
$\sim 10^9$\msun~at $z>3$ \citep[see, e.g., the results of the HST
CANDELS survey,][]{salmon15}. A different, complementary approach
involves exploiting gravitational lensing, i.e. using galaxy clusters
as natural telescopes able to amplify the light emitted by background
sources.  The HST Frontier Fields program \citep{lotz17} combined the
capabilities of the two photometric cameras ACS and WFC3 onboard HST
with the power of gravitational lensing to realize the deepest images
ever produced in six different pointings, each targeting one
$z\sim 0.3-0.5$ cluster together with their background galaxies. In
parallel, HST produced six images of close-by fields, referred to as
parallel fields.

In this paper we take advantage of the HST Frontier Fields program to
investigate the MS relation  at $1.3\leq z<6$ down to very low masses
($\log M/$\msun$\sim7.5$ at $z \lesssim 4$ and $\log M/$\msun$\sim 8$
at higher redshifts).  The paper is organized as
follows. Sect.~\ref{sec:data} describes the data set and the method
applied to estimate stellar masses and SFR.  Sects.~\ref{sec:ms}, \ref{sec:scatter} and
\ref{sec:ssfrevol} present our results, respectively on the MS
relation, its scatter and the evolution of the SFR per unit mass.
Finally, Sect.~\ref{sec:summ} summarizes the results.  In the
following, we adopt the $\Lambda$-CDM concordance cosmological model
(H$_0$ = 70 km/s/Mpc, $\Omega_M = 0.3$ and $\Omega_{\Lambda} = 0.7$)
and a \cite{salpeter55} IMF.  All magnitudes are in the AB system.

\section{Data set and method}\label{sec:data}

We use the multiwavelength catalogs of the first four HST Frontier
Fields clusters (field 1: Abell2744; field 2: MACS0416; field 3:
MACS0717; field 4: MACS1149) developed within the ASTRODEEP
project\footnote{ASTRODEEP is a coordinated and comprehensive program
  of i) algorithm/software development and testing; ii) data
  reduction/release; and iii) scientific data validation/analysis of
  the deepest multiwavelength cosmic surveys. For more information,
  visit http://astrodeep.eu}, publicly released by our
team\footnote{http://www.astrodeep.eu/frontier-fields/,\\http://www.astrodeep.eu/ff34/}
and also available through an interactive CDS
interface\footnote{http://astrodeep.u-strasbg.fr/ff/}.

The four fields are presented in \citet[fields 1 and 2]{merlin16} and
\citet[fields 3 and 4]{dicriscienzo17}. The 10 bands photometric
catalogs include the B435, V606 and I814 ACS/HST bands, the Y105,
J125, JH140 and H160 images from WFC3/HST, the K$_s$ band from
Hawk-I/VLT (fields 1 and 2) and MOSFIRE/Keck (fields 3 and 4) and IRAC
3.6 and 4.5\mic~data. The source detection has been carried out with
SExtractor \citep{bertin96} on the H band image, with the addition of
faint sources detected on a weighted average of the processed Y, J, JH
and H images. Photometry in the other bands has been derived with the
template-fitting code T-PHOT \citep{merlin15}, which uses galaxy
shapes in the detection band as prior information. The method adopted
to assemble the multiwavelength catalogs is described in details in
\cite{merlin16} and \cite{dicriscienzo17}. For our analysis we applied
a magnitude cut of H$<$27.5. This limit corresponds to a detection
completeness, assessed through simulations  taking into account
  the variation of the intracluster light across the image 
\citep{merlin16}, of 90-95\% for point-like sources and 50-80\% for
extended disks with 0.2'' half-light radius, depending on the field.

Photometric redshifts, when spectroscopy is unavailable (i.e. for 94\%
of the parent sample), have been obtained as the median value of six
different techniques, as detailed in \citet[fields 1 and
2]{castellano16} and \citet[fields 3 and 4]{dicriscienzo17}. This
approach reduces the scatter and the fraction of outliers as well as
minimizes systematics possibly associated to each individual method
\citep{dahlen13}.

Observed physical parameters, such as stellar masses, have been
derived with a SED fitting approach. We used \cite{bc03} stellar
population models and included nebular continuum and line emission
following \cite{schaerer09} as detailed in \cite{castellano14}. We
adopted delayed-$\tau$ star formation histories (SFH) of the form
$\psi(t) \propto t^2/\tau^3 \cdot {\rm exp}(-t/\tau)$. These are
rising-declining laws peaking at $t =2 \tau$. We set $\tau \geq 0.6$
Gyr, forcing high redshift galaxies ($z \gtrsim 4$) to experience the
increasing phase.  This choice is motivated by the results of
\cite{salmon15}, who measured on their high-z galaxy sample a SFH
increasing with time as a power-law with index 1.4 or maybe higher due
to possible incompleteness at low stellar mass. However, since stellar
masses are only mildly sensitive to the choice of the SFH
(\citealt{santini15}, although there may be exceptions for peculiar
galaxy populations, see also, e.g.,
\citealt{michalowski12,michalowski14}), this choice does not
significantly affect our results.  Stellar metallicities range from
0.02$Z_\odot$ to Solar, the dust reddening E(B-V) is comprised between
0 and 1.1 and the extinction law can be either \cite{calzetti00} or
SMC \citep{prevot84}.  1$\sigma$ uncertainties on the stellar masses
were computed by accounting for all the solutions within
$\chi_{min}+1$.

K and IRAC fluxes affected by heavy blending issues \citep{merlin15}
have not been considered in the SED fitting, neither for photo-z
\citep{castellano16} nor for physical
parameters. 
We checked that their removal has no systematic effect on the
  inferred stellar masses. We have then excluded from our analysis
sources at $z \geq 4$ whose K {\it and} IRAC fluxes have been ignored
as their SEDs result highly unconstrained and, as a consequence, their
inferred properties are highly unreliable. They amount to $\sim$7\% of
the H-selected sample.

SFRs were estimated from observed UV rest-frame photometry using the
same technique as \cite{castellano12}.  Briefly, the UV slope $\beta$
was obtained by fitting the observed photometric points adopting a
power-law approximation for the 1280--2600\AA~ spectral range, and the
\cite{meurer99} relation was assumed to infer the extinction used to
correct the 1600\AA~ luminosity. Finally, the dust-corrected
luminosity was converted into a SFR estimate with the
\cite{kennicutt12} factor.  This method avoids issues associated with
SED parameter degeneracy, particularly serious for the SFR estimate.
However, it limits the analysis to $z\geq 1.3$,
where the available photometry samples the UV rest-frame spectral
region. After visual inspection of sources with extreme values of the
UV slope, we removed sources with $\beta>1$
or $\beta
\leq -3.5$, mostly caused by noisy photometry
($\sim$8\%
of the sample in the redshift range analysed). 1$\sigma$
errors on the SFRs were computed by considering the uncertainty in
fitting the UV slope due to photometric errors as well as a scatter of
0.55 dex around the \cite{meurer99} relation \citep[see
also][]{fudamoto17}, with the two contributions summed in
quadrature. We note that recent ALMA results have found consistency
with the \cite{meurer99} law at $z\sim
3$, while suggesting a possible evolution towards lower attenuations
at $z$
larger than 4 or 5 \citep{capak15,bouwens16,fudamoto17}. The nature of
the attenuation law at high redshift is however still debated in the
recent literature, and consistency with a Calzetti/Meurer law is found
by some other studies \citep[e.g.][]{scoville15,cullen16}.

\begin{figure}[!t]
\center
\resizebox{\hsize}{!}{\includegraphics[angle=0]{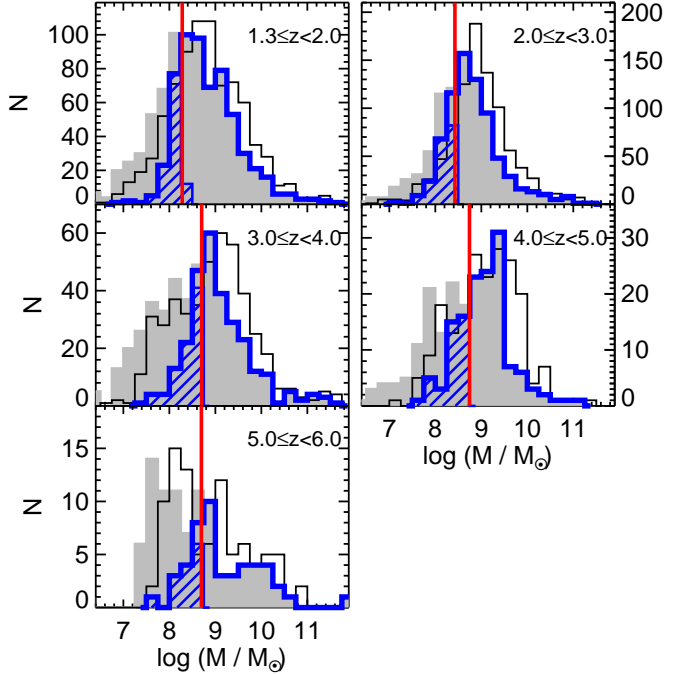}}
\caption{Stellar mass distribution for the $H<27.5$ sample in the four
  cluster fields in  different redshift bins.  The thin black
  open and grey shaded histograms show the observed and demagnified
  stellar mass distributions, respectively. The vertical red line
  shows the mass at which the sample is 90\% complete (see text). The
  thick blue open histograms represent the sample used in the
  analysis, with the hatched parts showing the galaxies recovered
  thanks to gravitational lensing.  These galaxies have stellar mass
  below the mass limit, but were included in the sample as their
  observed mass has been amplified above the threshold (see text).  }
\label{fig:masscompl}
\end{figure}

Intrinsic physical parameters were inferred after correcting for the
magnification due to gravitational lensing. The procedure adopted to
estimate the magnification factor from the shear and mass surface
density maps provided by different
teams\footnote{http://www.stsci.edu/hst/campaigns/frontier-fields/Lensing-Models}
is described in details in \cite{castellano16} (see \citealt{priewe17}
and \citealt{meneghetti17} for references and for a comparison
between different models). We adopted the median magnification, thus
excluding possible outlier values associated with a particular
model. For fields 1 and 2, at variance with the released catalogs,
 we only adopted the most updated maps available to date (i.e. v3).
 This results into the adoption of 6, 8, 7 and 7 models, respectively
 for the four cluster fields.  The magnification factor for the entire
 sample has a median value of 2, but can reach values as high as
 $\sim$60, as shown in \cite{castellano16}.

\begin{figure*}[!t]
\resizebox{\hsize}{!}{\includegraphics[angle=90]{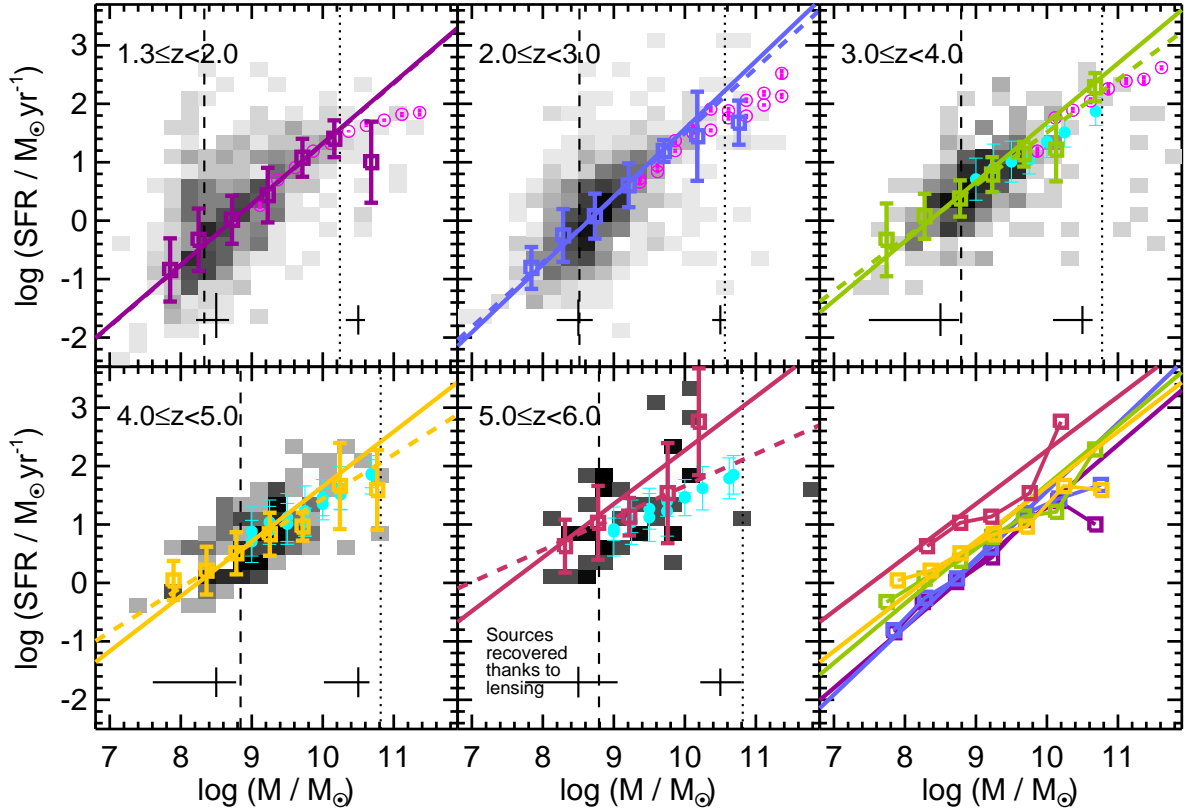}}
\caption{ Relation between the SFR and the stellar mass in different
  redshift bins.  The plane is coloured according to the density of
  sources, increasing from lightest (0.1\% of the total number of
  sources in that redshift interval) to darkest (6.3\%) shades on a
  linear scale. The open boxes show 2$\sigma$ clipped average
  values. The dashed coloured lines show the best-fit linear relation
  obtained with a 2$\sigma$ clipping procedure at stellar masses below
  the redshift-dependent turnover mass fitted by \cite{tomczak16}, to
  avoid the region where the linearity breaks up.  The solid coloured
  lines show the ``true'' underlying Main Sequence, after correcting
  for the Eddington bias. At the bottom of each redshift panel, the
  error bars show the median 1$\sigma$ uncertainties on stellar masses
  and SFRs for galaxies in the $\log M/$\msun$=$8-9 and 10-11
  bins. The dashed vertical lines show the observational mass
  completeness limit of the sample at the central redshift in each
  bin: all sources to the left of these lines have been recovered
  thanks to gravitational lensing. The dotted vertical lines show the
  turnover mass at the central redshift in each bin as taken from
  \cite{tomczak16}. At $z>4$ we adopted the threshold observed at
  $z=4$. The open magenta and solid cyan circles show the results of
  \cite{tomczak16} and \cite{salmon15}, respectively.  
  {\it
    Bottom right panel:} Average values and ``true'' Main Sequence
  relations in all redshift bins, colour-coded accordingly.}
\label{fig:ms}
\end{figure*}

\begin{deluxetable*}{ccc|ccc|ccc|ccc|ccc}[t!]
\tablecaption{SFR - Stellar mass relation average values for the cluster
  Frontier Fields. \label{tab:meanval}}  
 \tablehead{\multicolumn{3}{c}{$1.3\leq z<2$} & \multicolumn{3}{c}{$2\leq z<3$} &
  \multicolumn{3}{c}{$3\leq z<4$}
 & \multicolumn{3}{c}{$4\leq z<5$} & \multicolumn{3}{c}{$5\leq z<6$}\\
\colhead{(1)} & \colhead{(2)} & \colhead{(3)} & 
  \colhead{(1)} & \colhead{(2)} & \colhead{(3)} & \colhead{(1)} &
  \colhead{(2)} & \colhead{(3)} &  \colhead{(1)} & \colhead{(2)} &
  \colhead{(3)} & \colhead{(1)} & \colhead{(2)} & \colhead{(3)} 
}
\startdata
  7.85 & -0.84 & 0.54  &  7.84 & -0.81 & 0.36  &  7.74 & -0.33 & 0.62  &  7.90 & 0.04 & 0.33 & -- & -- & --\\
   8.25 & -0.33 & 0.53  &  8.29 & -0.25 & 0.45  &  8.28 &  0.07 & 0.39  &  8.36 &  0.21 & 0.41  &  8.31 &  0.63 & 0.45 \\
   8.72 &  0.02 & 0.41  &  8.73 &  0.08 & 0.39  &  8.78 &  0.38 & 0.31 &  8.78 &  0.51 & 0.36  &  8.78 &  1.03 & 0.63 \\
   9.23 &  0.44 & 0.47  &  9.21 &  0.61 & 0.37  &  9.23 &  0.79 & 0.30  &  9.25 &  0.83 & 0.37  &  9.22 &  1.13 & 0.32 \\
   9.71 &  1.08 & 0.32  &  9.70 &  1.19 & 0.19  &  9.68 &  1.14 & 0.21  &  9.72 &  0.96 & 0.24  &  9.76 &  1.54 & 0.86 \\
  10.16 &  1.40 & 0.31  & 10.17 &  1.44 & 0.76  & 10.13 &  1.22 & 0.54  & 10.24 &  1.66 & 0.74  & 10.20 &  2.76 & 0.91 \\
  10.68 &  1.00 & 0.69  & 10.76 &  1.68 & 0.38  & 10.68 &  2.28 & 0.24  & 10.77 &  1.59 & 0.68 & -- & -- & --
\enddata
\tablecomments{(1): $<$$\log M/$\msun$>$; (2): $<$$\log SFR/$(\msun/yr)$>$; (3):
  $\sigma_{\log SFR}$. Average values and associated scatter have been obtained with a
  2$\sigma$ clipping procedure.}
\end{deluxetable*}

\begin{deluxetable}{ccc}[t!]
\tablecaption{Main Sequence best-fit parameters (corrected for the
  Eddington bias). \label{tab:msparam}}  
\tablehead{
\colhead{Redshift} & \colhead{$\alpha$} & \colhead{$\beta$} 
}
\startdata
 $1.3\leq z<2$ & 1.04 $\pm$ 0.03 & 1.01 $\pm$ 0.04  \\
$2\leq z<3$   & 1.16 $\pm$ 0.03 & 1.22 $\pm$ 0.03\\
$3\leq z<4$   & 1.02 $\pm$ 0.04 & 1.37 $\pm$ 0.03\\
$4\leq z<5$   & 0.94 $\pm$ 0.06 & 1.37 $\pm$ 0.05 \\
$5\leq z<6$   & 0.92 $\pm$ 0.15 & 1.99 $\pm$ 0.13
\enddata
\tablecomments{Data have been fitted to the functional shape
  $\log SFR = \alpha \log (M/M_{9.7}) + \beta$, where $M_{9.7}=10^{9.7}$\msun, with a 2$\sigma$ clipping
  procedure. }
\end{deluxetable}

For studying the MS relation and inferring its slope, it is important
to consider a sample that is complete above a given stellar mass.  Our
sample is selected in the H band, but there is a dispersion in the
relation between H band magnitude and stellar mass.  Therefore, we
assessed the mass completeness limit of our sample by following the
procedure outlined in \cite{fontana04}. Briefly, considering a
passively evolving dust-free model, we compute the minimum mass given
the magnitude limit of the data and the $M/L$ distribution
distribution allowed by the adopted stellar library. Then we use the
observed $M/L$ distribution close to the magnitude limit at different
redshifts to account for the galaxies that are not observed.  We set
the mass limit as the mass above which 90\% of objects are observed.
This redshift-dependent limit is shown as a red vertical line in
Fig.~\ref{fig:masscompl}. However, a fraction of objects intrinsically below the
mass limit are actually observed above this threshold thanks to
gravitational lensing, boosting the flux from intrinsically fainter,
less massive objects. The magnification only depends on the relative
position of the lensed and lensing galaxies and not on their physical
properties.
In particular, it does not depend on their SFR, which will be randomly
distributed, in a statistical sense.  Based on this argument, we
included in our analysis galaxies with intrinsic stellar mass below
the mass limit, pushed above the threshold by gravitational lensing
magnification.  Fig.~\ref{fig:masscompl} shows how gravitational
lensing allows us to extend the analysis up to a factor of $\sim$10
below the strict observational mass limit (hatched region of the final
mass distribution represented by the blue histogram).  

After excluding 17 objects at $z>3$ which have been a-posteriori
visually inspected due to their suspect very high SFR with respect to
the MS, and which turned out to have bad fits due to very noisy
photometry, the final sample includes 1711 sources in the
redshift range $1.3\leq z < 6$.

\section{The Main Sequence relation}\label{sec:ms}

\subsection{Observed Main Sequence }\label{sec:msobs}

We show in Fig.~\ref{fig:ms} the relation between SFR and stellar
mass, i.e., the MS, in different redshift bins, from $z$$=$1.3
to $z$$=$6.
By means of a 2$\sigma$-clipping\footnote{We verified that this
  threshold is able to efficiently remove the MS outliers.} analysis,
we compute average values in bins of stellar mass (reported in
Table~\ref{tab:meanval}).  We recover the mass turnover at high
masses, at least out to $z$$=$3,
while above this value the turnover moves to higher masses where we
lack the required statistics due to the small sky area covered.  At
the opposite tail, we observe an extension of the MS relation down to
stellar masses only rarely probed in previous studies.

In the bottom right panel of Fig.~\ref{fig:ms} we report the average
SFR values in bins of stellar mass at different redshifts. We observe
an overall mild increase in the normalisation with redshift and a 
  mild flattening of the slope (at least out to $z\sim5$,
although we note that the highest redshift bin is noisier as it
includes fainter sources and is more affected by poor number
statistics).

\begin{figure*}[!t]
\resizebox{\hsize}{!}{
\includegraphics[angle=0]{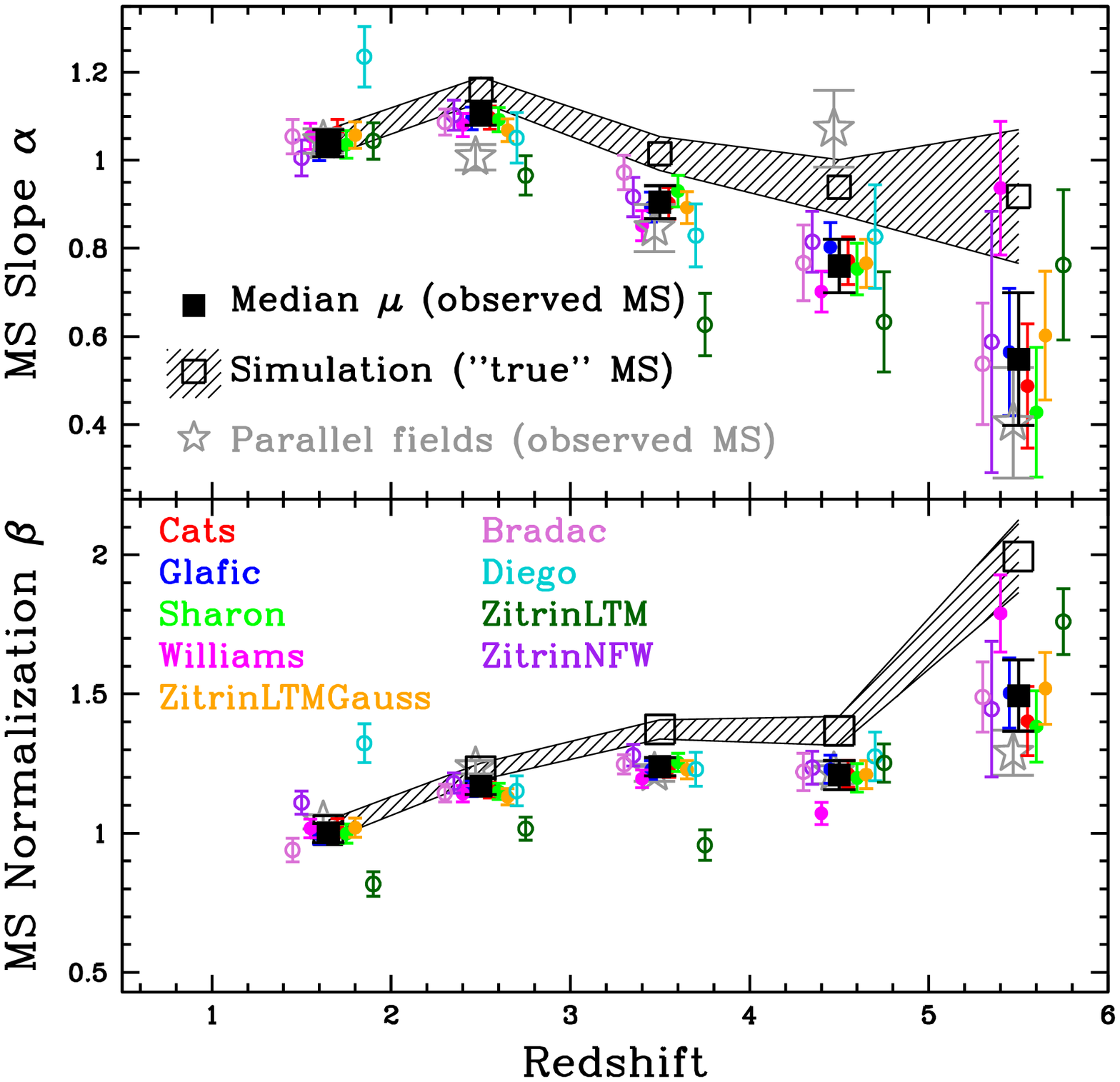}
\includegraphics[angle=0]{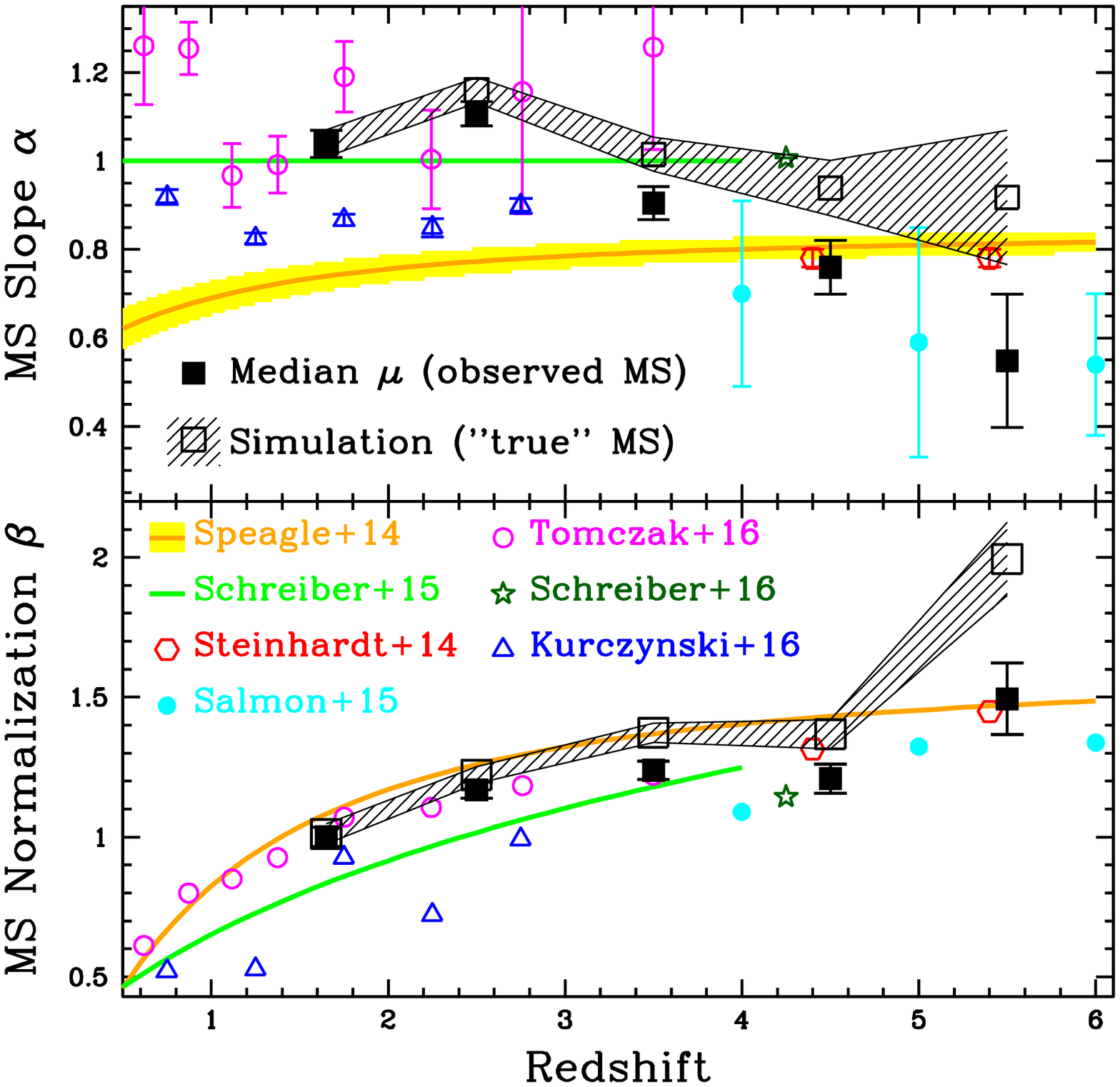}}
\caption{Evolution of the slope ({\it top panels}) and normalization at $\log
    M/M_\sun=9.7$ 
  ({\it bottom panels}) of the Main Sequence.  In both left and right panels, solid black squares show
  the observed MS parameters (computed adopting median magnification
  corrections), while open black
  squares and the hatched region show the MS parameters  and
    associated 1$\sigma$ uncertainties after correction for the
  Eddington bias (``true'' MS).  {\it Left panel:} 
 Coloured circles represent the observed MS parameters
  obtained using the different lensing models, as indicated by the
  legend in the bottom panel. Open coloured circles refer to lensing
  models that are not available for all fields:  only the fields
    for which the relevant model was available 
  have been used in these cases. As a consequence, these models
    show a 
    larger dispersion due to the
    decreased statistics (the results according to the `Diego' model,
    only available for field 2, have not been plotted in the highest
    redshift bin due to poor number statistics). For reference, the MS
    parameters obtained for the four parallel fields are also shown as
    grey open stars. The symbols are slightly shifted around the
  central redshift in each bin for visualization purposes. {\it
      Right panel:} Comparison with MS parameters inferred by previous
    works, according to the legend in the bottom panel. Normalizations have been rescaled
    to $\log M/M_\sun=9.7$, and the associated error bars have not been shown
    since they depend on the choice of the mass at which the MS is normalized.}  
\label{fig:lensing}
\end{figure*}

We fit our entire sample (not just the average values) 
with the linear relation 
\begin{equation}
\log SFR = \alpha \log (M/M_{9.7}) + \beta 
\end{equation}
where $M_{9.7}=10^{9.7}$\msun, avoiding the high-mass region where the
linearity breaks up. To determine this threshold we used the
redshift-dependent best-fit turnover mass obtained by \cite{tomczak16}
and its value at $z=4$ for higher redshift. We adopt a
2$\sigma$-clipping procedure to remove the outliers: we iteratively
reject all points more than 2$\sigma$ apart from the best-fit, where
$\sigma$ is calculated as the standard deviation of the residuals. The
best-fits are shown as dashed curves in Fig.~\ref{fig:ms}.  The
evolution of the MS slope and normalization is shown in
Fig.~\ref{fig:lensing} (black solid boxes). To verify the robustness
of our results against details associated with the magnification
correction, in the left panel of Fig.~\ref{fig:lensing} we also show
the best-fit MS parameters obtained with each individual lensing model
instead of the median magnification. The global picture is unaffected,
and the scatter among the various results can be regarded as a measure
of the observational uncertainty on the MS parameterization to be
associated with lensing modelling.  To further confirm that our
results are not driven by uncertainties in the magnification
corrections, we also compare the MS parameters with those obtained
from the parallel fields, where the magnification is very modest.

Overall, we apparently observe a decline in the slope from $\sim 1$ at $z<3$ to
$\sim0.6$ at $z\sim 6$, associated with an increase in the
normalization across the entire redshift range studied.

\subsection{Correction of the Eddington bias } \label{sec:simu}

Since measurement errors in the physical parameters can be high,
especially for high redshift faint sources, we tried to take into
account their influence on the inferred MS relation.  In other words,
we correct here the fitted MS for the Eddington bias, that is, we try
to recover the intrinsic, ``true'' MS, that, once convolved with
measurement errors, gives rise to the observed relation.

We run a Monte Carlo simulation comprising the following steps:
\begin{itemize}
\item Starting from an arbitrary MS, we populate the relation with a realistic logN-logS
  distribution and a given Gaussian scatter. To this aim, we start
  from the initial observed sample without applying any magnitude or
  mass cut. To increase the statistics we replicated the observed
  distribution 100 times. Given the observed redshift and stellar
  mass, we assign to each source the SFR predicted by the MS, starting
  from the observed relation and the observed scatter (the latter is
  plotted as a black line in Fig.~\ref{fig:scatter}). For stellar
  masses lower than $\log M/$\msun$=$8 and larger than $\log M/$\msun$=$10.3,
  where our measurement of the scatter is affected by poor number
  statistics, we adopt a standard value of 0.3 dex.
\item We randomly perturb each source in stellar mass and SFR adding
  Gaussian noise based on 
  the 1$\sigma$ uncertainties  on these parameters associated with each object.
\item We fit the resulting sample using exactly the same method
  adopted on real data, i.e., applying the magnitude and mass cuts and
  adopting a 2$\sigma$-clipping procedure to infer the MS. 
\item We iteratively modify the input MS, both in slope and normalization, until the best-fit simulated
  relation matches the observed MS  within a tolerance of 0.02 in both
   parameters. Similarly, we modify the input scatter
  until the simulated scatter is consistent the observed one.  At
    $\log M/M_\sun$$<$8 and  $\log M/M_\sun$$>$10.3 we assume an input
  scatter equal to the scatter in the 8$<$$\log M/M_\sun$$<$8.8 and
  9.6$<$$\log M/M_\sun$$<$10.3 mass bins, respectively, since our data prevent us to
  reliably evaluate the scatter outside this mass range. 
\item We iterate the procedure 100 times. The resulting parameters and
  associated uncertainties of the ``true'' MS as well as its
  scatter in bins of stellar mass are obtained by
  computing the average and standard deviation of the best-fit values
  and of the standard deviation of log(SFR)  
  over the 100 iterations.
\end{itemize}
The ``true'' MS are plotted as solid lines in Fig.~\ref{fig:ms}.  The
best-fit parameters are reported in Table~\ref{tab:msparam}, where the
uncertainties have been obtained by adding in quadrature the
uncertainties on the ``true'' MS with those on the observed MS. The
evolution of the ``true'' MS is shown by the hatched region in
Fig.~\ref{fig:lensing}.  Once corrected for the Eddington bias, the
slope of the MS is consistent with unity across the 1.3--6 redshift
range, while the normalization increases with redshift.

The Eddington bias has a negligible effect at $z \lesssim 3$, but it
is responsible for a flattening of the relation at higher redshift. It
is therefore crucial to take into account the effect of measure
uncertainties to reliably estimate the slope of the MS.

\subsection{Comparison with previous works}

As first step, we compare our results with the analysis of
\cite{tomczak16}, who measured the SFR by stacking on far-infrared
images.  As shown in Fig.~\ref{fig:ms}, we find excellent agreement
despite the completely different approach to estimate the SFR, both concerning the slope
and normalization of the relation as well as the turnover at high
masses. This agreement corroborates the use of dust-corrected UV
luminosity as a SFR tracer for the overall population.

At $z>3$ we compare our MS relations with those of \cite{salmon15},
based on the CANDELS survey, and once again the agreement is very good
(Fig.~\ref{fig:ms}), especially as far as the observed flattening at
high-$z$ is concerned. Compared to
CANDELS results, we managed, by exploiting gravitational lensing, to
extend the analysis to stellar masses more than a factor of 10 lower,
probing galaxies with $\log M/$\msun$\sim7.5$ at $z \lesssim 4$ and
$\log M/$\msun$\sim 8$ at higher redshifts.

We also compared our results with the deep analysis of
\cite{sawicki12}. They reach stellar masses as low as
$10^8$\msun~at
$z\sim
2.2$, although their sample is based on a UV-selection and is slightly
less complete towards dusty objects (in comparison, our sample is
$\sim
1.5$ magnitudes deeper in the UV and includes a few percent of source
with E(B-V)$>$0.5).
They find a comparable normalization and a flatter slope
($0.89\pm0.03$).

Another work of almost similar depth as ours is the analysis of
\cite{kurczynski16}, based on HUDF data. They reach stellar masses of
$10^7$ \msun~ at $z<2$ and $10^8$ \msun~ at $2<z<3$. They find roughly
constant slopes around 0.85-0.9 and increasing normalization with
decreasing cosmic times, both slightly lower than ours at comparable
redshifts.

We show in the right panel of Fig.~\ref{fig:lensing} a comparison of
the MS parameters reported by this work with those inferred by some of
the most recent studies as well as with the parameterization published
by \cite{speagle14} and based on a rich compilation of literature
results.  While there is a general consensus regarding an increasing
normalization with redshift
\citep[e.g.,][]{santini09,speagle14,schreiber15,tasca15,salmon15,lee15,tomczak16},
the slope is more uncertain and highly dependent on the details of the
analysis, on the SFR tracer adopted and on the sample selection
\citep{santini09,rodighiero14,speagle14}, as well as, as we showed, on
observational uncertainties.  Unevolving, approximately linear slopes,
as found by our analysis, are reported by the majority of studies
cited few lines above, especially at $z\lesssim 4$, while the high
redshift work of \cite{salmon15} based on CANDELS data reports
shallower slopes, ranging from $0.70\pm 0.21$ at $z\sim4$ to
$0.54 \pm 0.16$ at $z\sim 6$: 
their shallower slopes are probably a
consequence of the effect of observational uncertainties, that we have
corrected thanks to the simulation described in the
Sect.~\ref{sec:simu}.

\section{The scatter around the Main Sequence} \label{sec:scatter}

The existence of a relation on the SFR--stellar mass diagram as been
interpreted as evidence that galaxy growth is likely regulated by cold
gas accretion from the IGM on long timescales
\citep[e.g.][]{dekel09}. In this scenario, the tightness of the MS is
related to the level of similarity of the gas accretion histories
\citep[see e.g.][]{shimakawa17}. It is interesting to investigate 
whether the scatter around the MS depends on the stellar mass.

\begin{figure}[!t]
\resizebox{\hsize}{!}{\includegraphics[angle=90]{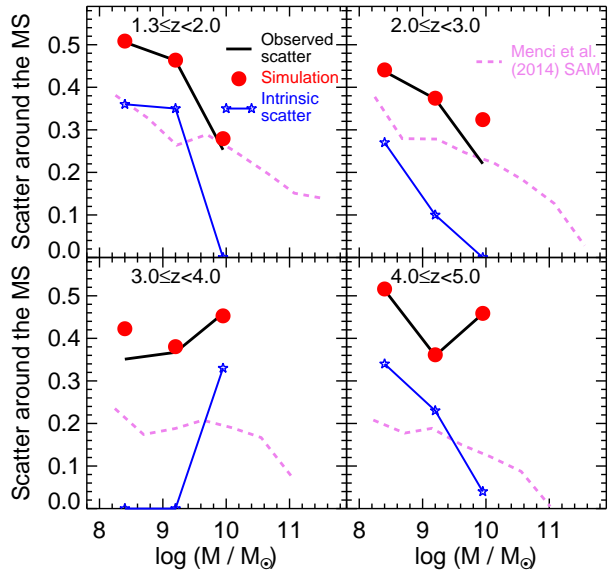}}
\caption{Scatter around the MS as a function of stellar mass, in
  different redshift bins. The black lines show the observed scatter,
  i.e. the standard deviation of log(SFR), as inferred through a
  2$\sigma$-clipping procedure. The red circles are the scatter
  computed from the simulation, using the same technique. Blue stars
  connected by blue lines show the intrinsic scatter, i.e. the scatter
  assumed as input in the simulation such that the output simulated
  scatter matches the observed one.  Dashed pink lines are the
  predictions of the hierarchical model of galaxy formation of
  \cite{menci14}, where the scatter has been computed with the same
  technique adopted on the data.  }
\label{fig:scatter}
\end{figure}

The observed scatter around the MS is computed as the standard
deviation of log(SFR) after a 2$\sigma$-clipping procedure to remove
the MS outliers, and it is shown by the black lines in
Fig.~\ref{fig:scatter} as a function of the stellar mass (we do not
show the highest redshift bin due to the higher noise level and poorer
statistics, which make it difficult to accurately evaluate the
scatter). If we focus on the common mass range
(9$\lesssim$log$M/$\msun$\lesssim$10), the observed scatter is
comparable with that measured by previous works
\citep[e.g.][]{salmon15,schreiber15}.

The intrinsic scatter around the MS is smaller than the observed one,
as the true scatter is convolved with the MS evolution within each bin
as well as with the scatter arising from uncertainties in deriving the
redshift and the physical parameters \citep{speagle14}. To account for
the effect of observational uncertainties, we consider the scatter
assumed as input to the simulation described in Sect.~\ref{sec:simu}
(blue stars and lines in Fig.~\ref{fig:scatter}), such that the
simulated scatter (red circles) matches the observed one.  This can be
regarded as the intrinsic scatter, except for the fact that it is
convolved with the evolution of the MS within each redshift bin. In
two of the redshift-mass bins the simulated scatter turns out to be
significantly larger than the observed one even by assuming no
intrinsic scatter in the simulation. This suggests that the large
observational uncertainties completely dominate the scatter of the
data points around the MS  and/or the simplified assumptions in
  our simulation (e.g. the assumption of Gaussian noise) hamper the
evaluation of the intrinsic scatter  in these bins. 

Overall however, we observe an indication, although not very robust, that the intrinsic
scatter decreases with the stellar mass, although this trend 
seems to be reversed (or is impossible to evaluate, as discussed
above) at $3<z<4$. This tentative trend is in contrast with recent
studies reporting no dependency on the stellar mass
\citep{whitaker12,schreiber15,salmon15,kurczynski16}, although not
extending to masses lower than $10^9$\msun (except for the latter
study). It instead agrees with the results of \cite{salim07}, who
finds that in the local Universe the scatter declines by 0.11 dex per
order of magnitude in stellar mass from $10^8$ to $10^{10.5}$\msun.

An increase in the scatter around the MS at low stellar masses as well
as low redshift is a natural outcome of theoretical models of galaxy
formation, and arises from the hierarchical structure formation and/or
from the highly efficient stellar feedback processes in small halos
\citep[e.g.][]{merlin12,lamastra13,hopkins14}.  To visualize this
effect, we plot in Fig.~\ref{fig:scatter} the predictions of the
semi-analytical model of \cite{menci14}, which, not surprisingly,
forecasts a smaller scatter than the observations as it is not
affected by measurement errors.  In a hierarchical scenario, high mass
galaxies have a large number of progenitors already collapsed at
high-$z$, as they formed in high density regions of the primordial
dark matter density field. The SFHs of these progenitor galaxies are
peaked at high redshift since the gas cooling and the star formation
efficiency are extremely efficient at early epochs. In addition, due
to the higher compactness and lower virial temperatures of high-$z$
dark matter halos, supernova feedback is less efficient at
heating/expelling the gas from the galaxy. Conversely, the smaller
number of progenitors of low mass galaxies, and their different
collapse time allowing also more prolonged star formation activity,
lead to a larger variety of their SFHs.

\section{The evolution of the specific SFR}\label{sec:ssfrevol}

We investigate here the evolution of the average, or typical, specific
SFR, i.e. the SFR per unit stellar mass (sSFR=SFR/$M$), which captures
information on the mass build-up process across cosmic time. Indeed,
theoretical models in which galaxy growth is dominated by cold accretion
predict that the sSFR should increase with redshift as $(1+z)^{2.25}$
\citep[e.g.][]{dekel09,dave11}.

Since the sSFR is not constant with the stellar mass (i.e. the slope
of the MS is not unitary at all redshifts), we derived the average
sSFR at constant stellar mass. We considered the
$\log M/$\msun$=9.5-10$ mass bin (average value $<\log
M/$\msun$>\sim
9.7$).
Despite adopting a fixed stellar mass at all redshifts implies
  considering a different galaxy population at different times,  
this allows an easy comparison with previous results computed at the
reference mass of $\log M/$\msun$=9.7$, and allows us to extend the
analysis of the evolution of the sSFR outside the redshift range
probed by the present work (i.e. at $z<1.3$ and $z>6$). 
We computed
the observed average values by means of a 2$\sigma$-clipping procedure
directly on the galaxies in this mass bin, without using the best-fit
MS relation (black solid squares). Given that the MS slope at high-$z$
is  sometimes found to be (slightly) sub-linear, the sSFR is a decreasing
function of the stellar mass. Not to run the risk of underestimating
the average sSFR, it is therefore essential to ensure completeness at
the chosen mass bin. The chosen mass bin is well above the completeness
level in our data.  Finally, to take into account the effect of observational
uncertainties, we make use of the Eddington-corrected MS estimated from the
simulation (Sect.~\ref{sec:simu}) and compute the typical sSFR at
$\log M/$\msun$=9.7$ (black open squares). We adopt the same procedure
for the data from the literature when the average value is not
available at the relevant mass.

\begin{figure}[!t]
\resizebox{\hsize}{!}{\includegraphics[angle=90,clip,viewport=140 0 720 830]{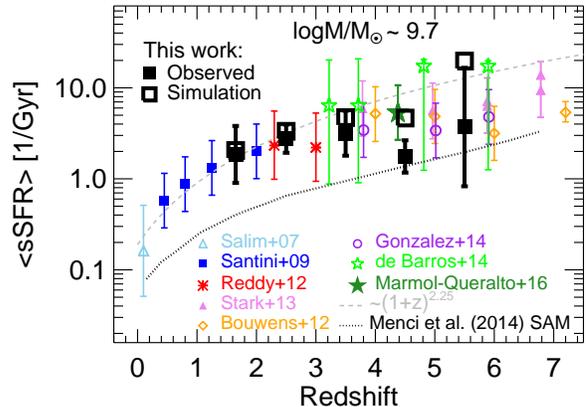}}
\caption{Evolution of the average sSFR as a function of redshift for
  $\log M/$\msun$\sim9.7$ galaxies.  Error bars show the standard
  deviation of log sSFR, computed with a 2-$\sigma$ clipping
  procedure.  Black squares are the results of this work: solid
  symbols are the observed values in the $\log M/$\msun$\sim9.5-10$
  mass bin, open ones are the result of the simulation, computed from
  the corrected MS at $\log M/$\msun$=9.7$. The other symbols present
  results from the literature according to the legend. When the
  average sSFR was not available at the chosen mass (as for
  \citealt{santini09} and \citealt{salim07}), we computed it by means
  of the best-fit MS relations.  As for the results of
  \cite{marmolqueralto16}, we only considered their 3.8-5.0 redshift bin, whose
  median stellar mass is equal to $\log M/$\msun$\sim9.7$, and we
  adopted a scatter of 0.3 dex. The black dotted line is the
  prediction of the semi-analytical model of galaxy formation of
  \cite{menci14} in the $\log M/$\msun$\sim9.5-10$
  mass bin. The dashed gray curve shows the prediction of
  accretion-dominated models \citep[e.g.][]{dekel09}, normalized to
  the bulk of data points at $z\sim2$. We note that the dip shown by
  our observed values at $z\sim 4.5$ is explained by the average lower
  SFR compared to the MS relation around the reference stellar mass
  because of poor number statistics (Fig.~\ref{fig:ms}). }
\label{fig:ssfrevol}
\end{figure}

Our results are in line with the compilation of data from previous
works (see Fig.~\ref{fig:ssfrevol}). Until a few years ago,
observations appeared to show that the sSFR increased steeply from
$z\sim0$ to $z\sim2$ \citep[e.g.][]{noeske07,santini09,karim11} and
flattened at higher redshift
\citep[e.g.][]{stark09,gonzalez10,reddy12,gonzalez10,labbe10,mclure11},
in sharp contrast with theoretical predictions \citep[see
e.g.][]{weinmann11}. The improvement in the dust-correction estimates
and the inclusion of nebular lines in SED modelling have led to higher
sSFR estimates at high redshift, alleviating the tension with
theoretical models. A moderate increase in the sSFR compared to its
value at $z\sim 2$ is now found in many studies
\citep[][]{stark13,gonzalez14,tasca15,marmolqueralto16,koprowski16},
especially at $z\gtrsim 5$. Nevertheless, the evolution seems overall
milder than that expected in an accretion-dominated scenario, although
some results are getting closer \citep{debarros14,faisst16}.  We note,
however, that given the scatter in the MS, a flat trend cannot be
ruled out.

We observe an evolution by a factor of $\sim 2$ from $z\sim 5.5$ to
$z\sim 1.6$.  Although a steeper trend is found when correcting for
the Eddington bias, the large scatter associated to the highest
redshift bin prevent us from drawing firm conclusions in this regard.
The observed decrease is lower than that predicted by theoretical
models in the same redshift interval (a factor of $\sim 8$), implying
that some sort of star formation suppression at high redshift is
required to match the observations (feedback effects can modify the
trend, see e.g. \citealt{dave11b}), that the effect of metallicity
should be taken into account \citep{krumholz12b} and that other
processes besides pure accretion are at play in assembling galaxies,
such as mergers \citep[e.g.][]{tasca15,faisst16}.  The evolution of
the sSFR is a critical observable to understand all these processes
regulating star formation and galaxy growth.  We note that a tension
between observations and theoretical models is also observed in terms
of absolute values of the sSFR (see the predictions of
\citealt{menci14} semi-analytical model). This effect is closely
connected with the well-known overall lower normalization of the
predicted MS compared to the observed one
\citep[e.g.][]{santini09,lamastra13}.

\section{Summary} \label{sec:summ}

We have studied the Main Sequence relation of star forming galaxies, i.e.,
the relation between galaxy SFR and stellar mass, in the first four HST Frontier
Fields.  Thanks to gravitational lensing amplification of faint
sources from massive foreground galaxy clusters, we could extend the
analysis to masses lower than has usually been possible with the
deepest data available before the Frontier Fields program and a factor
of $\sim$10 lower than most studies. We investigated the redshift
range $1.3\leq z< 6$ and probed stellar masses down to
$\log M/$\msun$\sim 7.5$ at $z\lesssim 4$ and $\log M/$\msun$\sim 8$
at higher redshifts. We find that the MS relation extends to such low masses. At the opposite side, we recover the mass
turnover at high masses at $z<3$, i.e., where we have enough
statistics at $\log M/$\msun$\gtrsim10.5$.  We run an accurate
Montecarlo simulation to take into account the effect of measure
uncertainties on stellar masses and SFRs and correct for the Eddington
bias. Such step is crucial to reliably measure the slope of the MS,
which tends to be flattened by observational errors at $z\gtrsim 3$.
We find increasing normalizations with redshift and an approximately
linear unevolving slope.

The combination of deep HST observations and gravitational lensing
allows us to observe tentative evidence that the scatter around the MS
relation increases at low stellar masses, although we cannot make a
strong claim.  If verified, this result implies a higher level of
uniformity in the SFHs at high masses, in agreement with observations
in the local Universe and with the predictions of theoretical models
of galaxy evolution.

We confirm the previously observed mild increase in the average sSFR
from $z\sim1.6$ to $z\sim5.5$ by a factor of $\sim$2, in tension with
the steeper trend predicted by accretion-driven models (a factor of
$\sim 8$ in the same redshift interval).  Being able to reproduce the
evolution of the sSFR is crucial to understand the processes
regulating the star formation and galaxy growth at different cosmic
epochs and assessing the relative importance of gas accretion versus
merging as well as the role of feedback mechanisms.

JWST observations in the next future will allow to extend the analysis
of the MS to even lower masses. Discerning between a MS that keeps
extending to faint sources and a broken relation will provide hints on
physical mechanisms such as feedback and reionization, or cold vs warm
dark matter scenarios.

\acknowledgments We thank the anonymous referee for the detailed
review and useful suggestions.  The research leading to these results
has received funding from the European Union Seventh Framework
Programme ASTRODEEP (FP7/2007-2013) under grant agreement n$^\circ$
312725. MJM acknowledges the support of the National Science Centre,
Poland through the POLONEZ grant 2015/19/P/ST9/04010. This project has
received funding from the European Union's Horizon 2020 research and
innovation programme under the Marie Sk{\l}odowska-Curie grant
agreement No.~665778.





\bibliographystyle{aasjournal}



\end{document}